# Evaluating research and researchers by the journal impact factor: is it better than coin flipping?

Ricardo Brito[a*], Alonso Rodríguez-Navarro[a,b]

[a] *Departamento de Estructura de la Materia, Física Térmica y Electrónica and GISC, Universidad Complutense de Madrid, Plaza de las Ciencias 3, 28040, Madrid, Spain*
[b] *Departamento de Biotecnología-Biología Vegetal, Universidad Politécnica de Madrid, Avenida Puerta de Hierro 2, 28040, Madrid, Spain*

*e-mail addresses*: RB, brito@ucm.es; AR-N, alonso.rodriguez@upm.es
\* Corresponding author: Ricardo Brito

The journal impact factor (JIF) is the average of the number of citations of the papers published in a journal, calculated according to a specific formula; it is extensively used for the evaluation of research and researchers. The method assumes that all papers in a journal have the same scientific merit, which is measured by the JIF of the publishing journal. This implies that the number of citations measures scientific merits but the JIF does not evaluate each individual paper by its own number of citations. Therefore, in the comparative evaluation of two papers, the use of the JIF implies a risk of failure, which occurs when a paper in the journal with the lower JIF is compared to another with fewer citations in the journal with the higher JIF. To quantify this risk of failure, this study calculates the failure probabilities, taking advantage of the lognormal distribution of citations. In two journals whose JIFs are ten-fold different, the failure probability is low. However, in most cases when two papers are compared, the JIFs of the journals are not so different. Then, the failure probability can be close to 0.5, which is equivalent to evaluating by coin flipping.

*Key words: research evaluation, impact factor, failure probability,*

## 1. Introduction

In a recent paper in *Nature World View*, John Tregoning (2018) upholds once more the opportunity of using the journal impact factor (JIF) to evaluate research or researchers: "The JIF is wrong in so many ways but it is so easy," the article highlights. Notably, the same can be said of coin flipping, which is not an argument to propose its use to decide which of two similar projects is funded or who of two similar researchers most merits an academic appointment. Although this comparison may sound as an exaggeration, we will show that failure probabilities almost as high as in coin flipping are in many cases associated with evaluations based on the JIFs.

The use of the JIF (Archambault & Larivière, 2009; Zhang, Rousseau, & Sivertsen, 2017) for research evaluations is extensive, alone or accompanied by other metrics (Hammarfelt & Rushforth, 2017), but it suffered an important setback in 2012 with the San Francisco Declaration on Research Assessment (DORA; https://sfdora.org/, accessed in 03/08/2018) and also because it is not used in the UK Research Excellence Framework (UK Forum for Responsible Reseaech Metrics, 2018), by the European Molecular Biology Organization (http://www.embo.org/documents/LTF/LTF_Guidelines_for_Applicants.pdf, accessed in 21/08/2018), and by other funding agencies. Furthermore, in 2016 the American Society of Microbiology reported that "the ASM Journals Editors in Chief and Society leadership have decided that Journal Impact Factors (JIFs) will no longer be posted to the journal websites or used in advertising. Our goal is to stop contributing further to the misuse of JIF as a proxy for evaluating the quality of an individual's scientific research" (https://www.asm.org/index.php/journals-impact-factor, accessed 06/08/2018).

Despite these notable rejections of the use of the JIF in research evaluation, the method is still in use, and opposite positions have been qualified and debated (Archambault & Larivière, 2009; Bornmann & Williams, 2017a; Bornmann & Williams, 2017b; Callaway, 2016; Hammarfelt & Rushforth, 2017; Peters, 2017; Seglen, 1997; Waltman & Traag, 2017). Regarding the high number of criticisms, Tregoning (2018) states: "But for all the invective heaped on the JIF as a metric, no alternative has emerged." Independently of the debate on the use of the JIF, this reasoning is not correct because if something were wrong, misleading, and inequitable the lack of an alternative is not a cause for continuing using it. The present study heaps more criticisms on the use of the JIF for evaluation purposes in comparison with the use of actual citation counts, but does not negate that publications in journals with high JIFs are associated with high scientific quality. Here a mathematical approach is applied to demonstrate that, with reference to actual citation counts, the use of the JIF to decide which of two papers has more merit for an academic appointment or project funding, in many cases, carries an inadmissible probability of taking a wrong decision.

The JIF "is calculated by dividing the number of current year citations to the source items published in that journal during the previous two years" (https://clarivate.com/essays/impact-factor/, accessed 08/14/2018). Thus, if the JIF is an average of the citations received by the papers published in a journal, it says nothing about the individual citations received by the papers that were taken to calculate the average; its use for the evaluation of individual papers has been called "ecological fallacy" (Leydesdorff, Wouters, & Bornmann, 2016). If the JIF is used as a surrogate of citation frequency a few papers will be correctly evaluated, when their number of citations coincides with the average, but most papers will be either over- or sub-evaluated because most papers will be below or above the average. The consequence is that when two papers in two journals are compared by using the JIFs as surrogates of citation frequencies, the result will be wrong when a JIF-sub-evaluated paper in a journal is compared with a JIF-over-evaluated paper in another journal. The flaw of the method is especially important because citation counts are well described by a lognormal distribution (Rodríguez-Navarro & Brito, 2018a and references therein), in which the most cited papers conform a heavy tail. Thus, the method leads to wrong and inequitable decisions, for which the probability of occurrence can be calculated.

Before the Tregoning's paper (2018) appeared, Waltman and Traag (2017) had raised an interesting debate about the validity of the JIF, finding that its use might not be wrong. The study uses two concepts, the value of an article (which is a "non observable concept") and either the number of citations or the JIF (which are "observable" concepts), and presents two alternative scenarios. In scenario 1, "the number of citations of an article is a *relatively accurate* indicator of the value of the article" (p. 13) and "journals are *rather heterogeneous* in terms of the values of the articles they publish" (p. 14). In scenario 2, "the number of citations of an article is a *relatively inaccurate* indicator of the value of the article" and "journals are fairly homogeneous in terms of the values of the articles they publish" (p. 14). In their study, Waltman and Traag (2017) provide different types of support for scenario 2.

For evaluations, the JIF is used in two different ways, assigning higher merit either to publications in journals with higher JIFs or to those that are in top percentiles of the lists of journals grouped in subject categories—for example, publications in Q1 journals (the top 25%; Bornmann & Marx, 2014; Bornmann & Williams, 2017a), versus publications in Q2 journals (within the top 25%–50% interval) in subject category lists of journals. In more extreme evaluations, only journals in the Q1 list are considered.

Considering two random papers in two journals with different JIFs, the present study calculates the probability that the random paper in the journal with the higher JIF is less cited than the random paper in the journal with the lower JIF. This probability is called here the failure probability because when this occurs the judgment by the number of citations contradicts the judgment by the JIF. With a different purpose, a similar approach has been used in a previous article, where the probability that a random paper in a journal receives more citations than a random paper in another journal is used to calculate the "citation success index" (Milojevic et al, 2017).

In this study, Section 3 reports the failure probabilities for research evaluations based on citation merits in three types of evaluations: by the JIF, the Q1 and Q2 quartiles, and publications in *Nature* and *Science*. In Section 4 we discuss the significance of our findings in real research evaluations, which should be aimed to evaluate the scientific merit.

As in most previous criticisms to the use of the JIF in research evaluations, this study assumes that journal articles are heterogeneous in terms of scientific relevance. However, its mathematical analysis is also valid to reject the use of the JIF as a proxy of the number of citations received by an article, independently of the interpretation that is given to this number.

## 2. Theoretical basis and dataset

*2.1. Failure probability*

To calculate the failure probabilities, we used the formula described in the Appendix, assuming that the number of citations of journal papers follow a lognormal distribution, for which there is strong support (Rodríguez-Navarro & Brito, 2018a and references therein); a lognormal distribution for monodisciplinary journals has also been specifically investigated (Thelwall, 2016). The $\mu$ and $\sigma$ parameters of the lognormal functions were calculated as the mean and standard deviation of the log transformed numbers of citations, omitting the publications with zero citations. We prefer this approach to that of adding 1 to all citations because it was found that in some journals the number of uncited articles was unexpectedly high. This especially occurs when journals include letters and comments, which frequently do not receive citations. The mix of these probably uncited publications with the statistically predetermined uncitable research articles (Thelwall, 2016) increases the number of zero citations. Considering this problem, this study does not include journals with a high number of uncited papers to avoid a probable bias, since uncited papers are included in the calculation of the JIF but not in this paper's calculations. Furthermore, publications that are not or seldom cited because they are not real research publications decrease the goodness of fit of the lognormal distribution of regular research publications.

To ascertain that the goodness of fit to a lognormal distribution of the journals included in the study was high, the mean number of citations calculated from the actual data was compared with the mean calculated from the $\mu$ and $\sigma$ parameters by the formula

$$m = \exp(\mu + \sigma^2/2) \tag{1}$$

In many cases the difference between the two referred means is small—less than 2%—but in others the difference is high. The most important reason for high deviations of the means calculated in these two ways is the publication of review articles together with original articles. This practice is so frequent that it constituted a real problem in the selection of high-JIF journals for this study. For example, in *Current Biology* and *Plant Cell* the deviations of the two means were 24.5% and 23.7%, respectively, considering "all publications" but only 1.9 and 2.6%, respectively, when only "articles" were considered. In order to avoid an excessive restriction of journals, especially in Q1, a limit of a 6% deviation was fixed for the journals included in this study, which supposed a small deviation from the lognormal distribution.

It is worth noting that the mean number of citations reported in this study should be highly correlated with the JIF, but is not the JIF multiplied by four (in our case the citation window is four years) because the mean and the JIF are calculated from different citation years.

*2.2. Searches and dataset*

All citation data were obtained from the *Web of Science*; journal lists by categories and JIFs were obtained from the *Journal Citation Reports* (*JCR* categories, which group journals with similar research subjects). The number of paper citations in each year was obtained by using the Create Citation Report tool of the database.

To perform this study, we selected a collection of journals in natural sciences and technology. First, we selected a list of JCR categories in which the JIF values of the first journals varied from high (e.g., Biochemistry & Molecular Biology) to low (e.g., Engineering Multidisciplinary). Next, we eliminated the categories with a low number of journals (< 100) and selected six categories that covered a wide range of JIF values (Biochemistry & Molecular Biology, Immunology, Neurosciences, Physics Applied,

Environmental Sciences, and Engineering, Electrical & Electronic). In these categories, we selected six journals, three in Q1 and three in Q2, homogeneously distributed over the range of the JIF values.

We studied publications in 2012, recording the JIF in 2012 and the number of citations that these papers received in 2014–2017. This medium-term number of citations eliminating the first year was used to avoid the variability that is associated with the number of citations in the first year after publication and with too-short citation windows. Furthermore, it is known than one or two years after publication the JIFs (Abramo, D'Angelo, & Di Costa, 2010) or combined bibliometric indicators based on the JIFs (Levitt & Thelwall, 2011) offer useful information. Although publications in 2012 are not considered in the calculation of the 2012 JIF, this JIF was used because in Spain, and probably in other countries, the JIF considered for the evaluation of a paper is that corresponding to the paper's publication year; the same decision is made in academic studies (Levitt & Thelwall, 2011). The database was accessed in July 2018.

## 3. Results

Describing the probability of failing of JIF research evaluation as the probability that the judgments by the JIF and citation counts are contradictories (Section 1), it seems that the use of the JIF for judging papers implies a high probability of failing. This occurs because the citation distributions of two journals that have different impact factors overlaps at low numbers of citations (Larivière et al, 2016), especially if the JIFs are not very different. Figure 1 illustrates this fact presenting the distribution of citations in 2014–2017 to 2012 papers in *Water Research* and *Chemosphere*. In 2012, the JIFs of these journals were 4.66 and 3.14, and they published 650 and 752 papers, respectively. The histogram suggests that papers with fewer than 20 citations are similarly probable in both journals. Papers with a higher number of citations are more probable in *Water Research* than in *Chemosphere*, but still it can be guessed that taking a paper at random from each journal, it is not improbable that the paper in *Chemosphere* is the one that receives more citations; in these cases, the evaluation by the JIF as a surrogate of citation frequency will be wrong.

*3.1. Probability of failure in evaluations based on the JIF*

On the basis that evaluating by the JIF implies a certain probability of assigning higher citation merit to the paper that has the lower number of citations, the research question is how wide the difference between two JIFs has to be in order that the probability of failure be low enough to make failure unlikely. This probability of failure depends on both the $\mu$ and $\sigma$ parameters of the citation distributions of the two journals under consideration (Appendix). These parameters are highly variable among journals, and their link with the JIF is complex as commented in Section 2.1. Therefore, the relationship between the two JIFs involved and the probabilities of failing must be tackled empirically. For this purpose we selected 39 journals that fulfill the conditions described in Section 2.1; Table 1 shows the list of journals that were selected, which had JIFs ranging from 14.8 to 1.3. As could be expected, the JIFs of these journals in 2012 are highly correlated with the mean number of citations in 2014–2017 of the papers published in 2012 (Pearson correlation coefficient = 0.91, $p < 10^{-15}$) but showed large individual deviations. Therefore, if the journals are ordered by the JIF, they are not ordered by the mean. For example, the mean of *Water Research* (journal #11 if ordered by the JIF) is three times higher than that of Clinical and Experimental Allergy (journal #10 if ordered by the JIF).

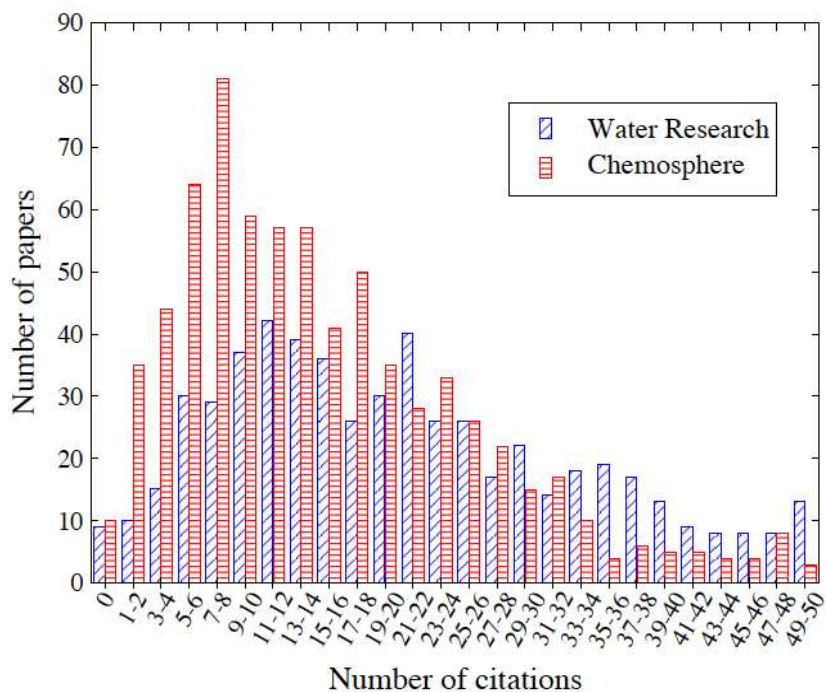

*Figure 1. Distribution of the citations to papers in the journals Water Research and Chemosphere. Citations in 2014–2017 to papers published in 2012*

Next, the matrix of failure probabilities resulting from pairwise comparisons of these journals, which had been ordered from higher to lower JIFs, was constructed (Table 2; the matrix is large and has been divided by odd and even journal numbers for printing reasons; the complete matrix is presented as supplementary material, Table S1). The most apparent characteristic of this matrix is that many probability values are very close to 0.5. In most cases, these high probability values correspond to journals with similar JIFs—they are close to the matrix diagonal—, but this is not a perfect rule because, as mentioned earlier, the journals when ordered by their JIFs are not ordered according to the mathematical properties of their lognormal distributions. For this same reason a few probabilities are higher than 0.5; for example, journals #5 versus journal #11 is 0.68.

Table 1. Journals that have been used to calculate the failure probability when real merits are calculated by citation counting and evaluation merits are assigned according to the Journal Impact Factor (JIF)[a]

| # | Journal title | Papers | JIF | μ | σ | Mean |
|---|---|---|---|---|---|---|
| 1 | Advanced Materials | 893 | 14.83 | 3.75 | 1.10 | 19.07 |
| 2 | Genome Research | 246 | 14.40 | 3.71 | 1.10 | 18.62 |
| 3 | Neuropsychopharmacology | 312 | 8.68 | 2.88 | 1.00 | 6.10 |
| 4 | Cell Death and Differentiation | 210 | 8.37 | 2.96 | 0.94 | 7.45 |
| 5 | Biomaterials | 905 | 7.60 | 3.38 | 0.84 | 10.21 |
| 6 | Molecular Ecology | 476 | 6.28 | 2.74 | 1.07 | 6.60 |
| 7 | Journal of Infectious Diseases | 629 | 5.85 | 2.47 | 1.09 | 4.47 |
| 8 | Neurobiology of Diseases | 303 | 5.62 | 2.77 | 0.85 | 5.44 |
| 9 | Molecular Microbiology | 359 | 4.96 | 2.62 | 0.81 | 4.44 |
| 10 | Clinical and Experimental Allergy | 338 | 4.79 | 2.32 | 1.10 | 2.38 |
| 11 | Water Research | 650 | 4.66 | 3.05 | 0.87 | 7.41 |
| 12 | Experimental Neurology | 362 | 4.65 | 2.63 | 0.98 | 5.12 |
| 13 | Journal of Molecular Biology | 439 | 3.90 | 2.35 | 0.90 | 3.59 |
| 14 | European Journal of Neuroscience | 368 | 3.75 | 2.30 | 0.92 | 3.54 |
| 15 | Pediatric Infectious Disease Journal | 334 | 3.57 | 1.83 | 0.97 | 2.16 |
| 16 | Plant Molecular Biology | 133 | 3.52 | 2.51 | 0.90 | 4.26 |
| 17 | Neuropsychologia | 389 | 3.48 | 2.43 | 0.82 | 3.82 |
| 18 | IEEE Transactions on Geoscience and Remote Sensing | 417 | 3.47 | 2.52 | 1.10 | 5.32 |
| 19 | Clinical Experimental Immunology | 182 | 3.41 | 2.17 | 0.87 | 2.99 |
| 20 | BMC Cancer | 620 | 3.33 | 2.39 | 0.88 | 3.99 |
| 21 | Science of the Total Environment | 1024 | 3.26 | 2.61 | 0.88 | 5.06 |
| 22 | Journal of Inorganic Biochemistry | 265 | 3.20 | 2.30 | 0.90 | 3.58 |
| 23 | Clinical Neurophysiology | 345 | 3.14 | 2.04 | 1.05 | 0.50 |
| 24 | Chemosphere | 752 | 3.14 | 2.55 | 0.88 | 4.57 |
| 25 | Phytochemistry | 249 | 3.05 | 2.25 | 0.78 | 3.20 |
| 26 | Journal of Neuroimmunology | 645 | 3.03 | 1.98 | 0.90 | 0.66 |
| 27 | IEEE Transactions on Information Theory | 504 | 2.62 | 2.11 | 1.12 | 3.78 |
| 28 | Analytical Biochemistry | 441 | 2.58 | 1.79 | 0.93 | 2.21 |
| 29 | Marine Pollution Bulletin | 563 | 2.53 | 2.38 | 0.92 | 2.85 |
| 30 | Ecotoxicology and Environmental Safety | 372 | 2.20 | 2.36 | 0.84 | 3.63 |
| 31 | American Mineralogist | 215 | 2.20 | 1.86 | 0.94 | 2.36 |
| 32 | IEEE Photonic Technology Letters | 684 | 2.04 | 1.78 | 1.00 | 2.15 |
| 33 | Environmental Toxicology and Pharmacology | 170 | 2.01 | 1.88 | 0.91 | 2.40 |
| 34 | Journal of the Science of Food and Agriculture | 418 | 1.76 | 1.88 | 0.87 | 2.25 |
| 35 | IEEE Transactions on Industry Applications | 227 | 1.67 | 2.43 | 1.10 | 4.72 |
| 36 | Engineering Applications of Artificial Intelligence | 157 | 1.63 | 2.15 | 1.02 | 3.58 |
| 37 | IEER transactions on Magnetics | 1013 | 1.42 | 1.64 | 0.99 | 1.87 |
| 38 | Journal of Mathematical Physics | 513 | 1.30 | 1.30 | 0.89 | 1.11 |
| 39 | Journal of Vacuum Science & Technology B | 334 | 1.27 | 1.39 | 0.91 | 1.29 |

[a] JIF in 2012. Citations in 2014-2017 to publications in 2012. The values of μ and σ have been calculated based on a lognormal distribution. Mean number of citations in 2014-2017.

The general conclusion that can be drawn from the data recorded in Table 2 and Table S1 is that in many cases the failure probability of using the JIF for the evaluative comparison of the citation merits of two papers is quite high (≈ 0.5–0.3). Only if the difference between the JIFs is very high—for example, 14.8 (journal #1) and 1.3 (journal #39)—is the failure probability really low (≈0.05).

The same general conclusion can be reached from Table 2 in (Milojevic et al, 2017). The figures in our study are similar but not identical to those previously published because the methods of calculation are different.

*3.2. Probability of failure in evaluations based on the quartile position of journals*

Some evaluations are performed by prioritizing papers published in Q1 journals (top 25%) versus those published in Q2 journals (within the top 25%–50% interval) in the same *JCR* category. Because it is obvious that the JIFs of the journals in the lower part of the Q1 set and in the upper part of the Q2 set will be very similar, the findings in the previous section (Table 2) suggest that the failure probability in evaluations by the Q1 indicator can be very close to 0.5. Furthermore, the data in Table 2 suggest that even in the comparison between journals in the upper part of Q1 and in the lower part of Q2, the expected differences in the JIFs will not be high enough as to predict low probabilities of failing.

To calculate the probabilities of failing in evaluations based on journal quartiles, we considered six journals in four *JCR* categories: Neurosciences, Biochemistry & Molecular Biology, Environmental Sciences, and Engineering, Electrical & Electronic. Three out of the six journals were distributed in Q1 and the other three were distributed in Q2. Table 3 shows the probability matrices of pairwise comparisons of the six journals in each *JCR* category. As expected, the lower probabilities of failure occurred when comparing the two journals with the highest and lowest JIFs, the first in Q1 with the last in Q2. However, even in this case the probability of failure was high—approximately 0.2.

Table 2. Failure probability when real merits are calculated by citation counting and evaluating merits are assigned according to the Journal Impact Factor (JIF)[a]

| # | 1 | 3 | 5 | 7 | 9 | 11 | 13 | 15 | 17 | 19 | 21 | 23 | 25 | 27 | 29 | 31 | 33 | 35 | 37 | 39 |
|---|---|---|---|---|---|---|---|---|---|---|---|---|---|---|---|---|---|---|---|---|
| 1 | 0.50 | 0.28 | 0.39 | 0.20 | 0.20 | 0.31 | 0.16 | 0.10 | 0.17 | 0.13 | 0.21 | 0.13 | 0.13 | 0.15 | 0.17 | 0.10 | 0.10 | 0.20 | 0.08 | 0.05 |
| 3 | | 0.50 | 0.65 | 0.39 | 0.42 | 0.55 | 0.35 | 0.23 | 0.36 | 0.30 | 0.42 | 0.28 | 0.31 | 0.30 | 0.36 | 0.23 | 0.23 | 0.38 | 0.19 | 0.14 |
| 5 | | | 0.50 | 0.25 | 0.26 | 0.39 | 0.20 | 0.11 | 0.21 | 0.16 | 0.26 | 0.16 | 0.16 | 0.18 | 0.21 | 0.11 | 0.11 | 0.25 | 0.09 | 0.05 |
| 7 | | | | 0.50 | 0.54 | 0.66 | 0.47 | 0.33 | 0.49 | 0.41 | 0.54 | 0.39 | 0.43 | 0.41 | 0.47 | 0.34 | 0.34 | 0.49 | 0.29 | 0.22 |
| 9 | | | | | 0.50 | 0.64 | 0.41 | 0.27 | 0.43 | 0.35 | 0.50 | 0.33 | 0.37 | 0.36 | 0.42 | 0.27 | 0.27 | 0.44 | 0.22 | 0.16 |
| 11 | | | | | | 0.50 | 0.29 | 0.17 | 0.30 | 0.24 | 0.36 | 0.23 | 0.25 | 0.25 | 0.30 | 0.18 | 0.18 | 0.33 | 0.14 | 0.09 |
| 13 | | | | | | | 0.50 | 0.35 | 0.53 | 0.44 | 0.58 | 0.41 | 0.47 | 0.43 | 0.51 | 0.35 | 0.36 | 0.52 | 0.30 | 0.23 |
| 15 | | | | | | | | 0.50 | 0.68 | 0.60 | 0.72 | 0.56 | 0.63 | 0.57 | 0.66 | 0.51 | 0.51 | 0.66 | 0.45 | 0.37 |
| 17 | | | | | | | | | 0.50 | 0.41 | 0.56 | 0.38 | 0.44 | 0.41 | 0.48 | 0.32 | 0.33 | 0.50 | 0.27 | 0.20 |
| 19 | | | | | | | | | | 0.50 | 0.64 | 0.46 | 0.53 | 0.48 | 0.57 | 0.40 | 0.41 | 0.57 | 0.34 | 0.27 |
| 21 | | | | | | | | | | | 0.50 | 0.34 | 0.38 | 0.36 | 0.43 | 0.28 | 0.28 | 0.45 | 0.23 | 0.17 |
| 23 | | | | | | | | | | | | 0.50 | 0.56 | 0.52 | 0.60 | 0.45 | 0.45 | 0.60 | 0.39 | 0.32 |
| 25 | | | | | | | | | | | | | 0.50 | 0.46 | 0.54 | 0.37 | 0.38 | 0.55 | 0.31 | 0.24 |
| 27 | | | | | | | | | | | | | | 0.50 | 0.57 | 0.43 | 0.44 | 0.58 | 0.38 | 0.31 |
| 29 | | | | | | | | | | | | | | | 0.50 | 0.35 | 0.35 | 0.51 | 0.29 | 0.22 |
| 31 | | | | | | | | | | | | | | | | 0.50 | 0.51 | 0.65 | 0.44 | 0.36 |
| 33 | | | | | | | | | | | | | | | | | 0.50 | 0.65 | 0.43 | 0.35 |
| 35 | | | | | | | | | | | | | | | | | | 0.50 | 0.30 | 0.23 |
| 37 | | | | | | | | | | | | | | | | | | | 0.50 | 0.43 |
| 39 | | | | | | | | | | | | | | | | | | | | 0.50 |

[a] The name of the journals, JIF, and parameters $\mu$ and $\sigma$ of the lognormal distribution of citations of each journal are reported in Table I. Probabilities have been calculated as described in text.

Table 2 (continued)

| # | 2 | 4 | 6 | 8 | 10 | 12 | 14 | 16 | 18 | 20 | 22 | 24 | 26 | 28 | 30 | 32 | 34 | 36 | 38 |
|---|---|---|---|---|----|----|----|----|----|----|----|----|----|----|----|----|----|----|----|
| 2 | 0.50 | 0.30 | 0.26 | 0.25 | 0.19 | 0.23 | 0.16 | 0.20 | 0.22 | 0.17 | 0.16 | 0.20 | 0.11 | 0.09 | 0.16 | 0.10 | 0.10 | 0.15 | 0.04 |
| 4 | | 0.50 | 0.44 | 0.44 | 0.33 | 0.40 | 0.31 | 0.36 | 0.38 | 0.33 | 0.31 | 0.38 | 0.23 | 0.19 | 0.32 | 0.19 | 0.20 | 0.28 | 0.10 |
| 6 | | | 0.50 | 0.51 | 0.39 | 0.47 | 0.38 | 0.43 | 0.44 | 0.40 | 0.38 | 0.45 | 0.29 | 0.25 | 0.39 | 0.26 | 0.27 | 0.34 | 0.15 |
| 8 | | | | 0.50 | 0.37 | 0.46 | 0.35 | 0.42 | 0.43 | 0.38 | 0.35 | 0.43 | 0.26 | 0.22 | 0.37 | 0.23 | 0.23 | 0.32 | 0.12 |
| 10 | | | | | 0.50 | 0.58 | 0.49 | 0.55 | 0.55 | 0.52 | 0.49 | 0.56 | 0.41 | 0.36 | 0.51 | 0.36 | 0.38 | 0.45 | 0.24 |
| 12 | | | | | | 0.50 | 0.40 | 0.46 | 0.47 | 0.43 | 0.40 | 0.48 | 0.31 | 0.27 | 0.42 | 0.27 | 0.28 | 0.37 | 0.16 |
| 14 | | | | | | | 0.50 | 0.56 | 0.56 | 0.53 | 0.50 | 0.58 | 0.40 | 0.35 | 0.52 | 0.35 | 0.37 | 0.46 | 0.22 |
| 16 | | | | | | | | 0.50 | 0.50 | 0.46 | 0.43 | 0.51 | 0.34 | 0.29 | 0.45 | 0.29 | 0.31 | 0.40 | 0.17 |
| 18 | | | | | | | | | 0.50 | 0.46 | 0.44 | 0.51 | 0.35 | 0.31 | 0.45 | 0.31 | 0.32 | 0.40 | 0.19 |
| 20 | | | | | | | | | | 0.50 | 0.47 | 0.55 | 0.37 | 0.32 | 0.49 | 0.32 | 0.34 | 0.43 | 0.19 |
| 22 | | | | | | | | | | | 0.50 | 0.58 | 0.40 | 0.35 | 0.52 | 0.35 | 0.37 | 0.46 | 0.21 |
| 24 | | | | | | | | | | | | 0.50 | 0.33 | 0.28 | 0.44 | 0.28 | 0.29 | 0.38 | 0.16 |
| 26 | | | | | | | | | | | | | 0.50 | 0.44 | 0.62 | 0.44 | 0.47 | 0.55 | 0.30 |
| 28 | | | | | | | | | | | | | | 0.50 | 0.68 | 0.50 | 0.53 | 0.60 | 0.35 |
| 30 | | | | | | | | | | | | | | | 0.50 | 0.33 | 0.35 | 0.44 | 0.19 |
| 32 | | | | | | | | | | | | | | | | 0.50 | 0.53 | 0.60 | 0.36 |
| 34 | | | | | | | | | | | | | | | | | 0.50 | 0.58 | 0.32 |
| 36 | | | | | | | | | | | | | | | | | | 0.50 | 0.27 |
| 38 | | | | | | | | | | | | | | | | | | | 0.50 |

Table 3. Failure probability when real merits are calculated by citation counting and evaluating merits are assigned according to the rank of the journal in Q1 (top 25%) and Q2 (within the top 25% and 50%) in the category lists of the Journal of Citations Reports[a]

| #journal[b] | JIF | #category[c] | 1 | 2 | 3 | 4 | 5 | 6 |
|---|---|---|---|---|---|---|---|---|
| Neurosciences | | | | | | | | |
| 3 | 8.68 | 1 | 0.50 | 0.47 | 0.43 | 0.33 | 0.36 | 0.28 |
| 8 | 5.62 | 2 | | 0.50 | 0.46 | 0.35 | 0.39 | 0.29 |
| 12 | 4,65 | 3 | | | 0.50 | 0.40 | 0.44 | 0.34 |
| 14 | 3.75 | 4 | | | | 0.50 | 0.54 | 0.43 |
| 17 | 3.48 | 5 | | | | | 0.50 | 0.38 |
| 23 | 3.14 | 6 | | | | | | 0.50 |
| Biochemistry & Molecular Biology | | | | | | | | |
| 4 | 8.37 | 1 | 0.50 | 0.44 | 0.39 | 0.32 | 0.31 | 0.28 |
| 6 | 6.28 | 2 | | 0.50 | 0.46 | 0.39 | 0.38 | 0.36 |
| 9 | 4.96 | 3 | | | 0.50 | 0.41 | 0.40 | 0.37 |
| 13 | 3.90 | 4 | | | | 0.50 | 0.48 | 0.47 |
| 22 | 3.20 | 5 | | | | | 0.50 | 0.48 |
| 25 | 3.05 | 6 | | | | | | 0.50 |
| Environmental Sciences | | | | | | | | |
| 11 | 4.66 | 1 | 0.50 | 0.36 | 0.34 | 0.30 | 0.28 | 0.18 |
| 21 | 3.26 | 2 | | 0.50 | 0.48 | 0.43 | 0.42 | 0.28 |
| 24 | 3.14 | 3 | | | 0.50 | 0.45 | 0.44 | 0.30 |
| 29 | 2.53 | 4 | | | | 0.50 | 0.49 | 0.35 |
| 30 | 2.20 | 5 | | | | | 0.50 | 0.35 |
| 33 | 2.01 | 6 | | | | | | 0.50 |
| Engineering, Electrical & Electronic | | | | | | | | |
| 18 | 3.47 | 1 | 0.50 | 0.40 | 0.31 | 0.48 | 0.28 | 0.21 |
| 27 | 2.62 | 2 | | 0.50 | 0.41 | 0.58 | 0.38 | 0.31 |
| 32 | 2.04 | 3 | | | 0.50 | 0.67 | 0.46 | 0.39 |
| 35 | 1.67 | 4 | | | | 0.50 | 0.30 | 0.23 |
| 37 | 1.42 | 5 | | | | | 0.50 | 0.43 |
| 39 | 1.27 | 6 | | | | | | 0.50 |

[a] Probabilities have been calculated as described in text
[b] The name, ranking number, and parameters of the lognormal distributions are recorded in Table 1.
[c] The ranking number in category is the following: 1, 2, 3, and 4, 5, 6 correspond to journals in the upper, medium, and lower positions in the Q1 and Q2 sets of journals, respectively.

### 3.3. Probability of failure in evaluations based on publications in Nature and Science

Currently, researchers suffer great pressure to publish in some specific, highly prestigious journals (Lawrence, 2003), and among these journals, *Nature* and *Science* are the most valued. From a mathematical point of view, this method uses the same basis as evaluations by the JIF; i.e., all publications in these selected journals have the same high merit and it does not matter whether the method invokes the high JIFs of these journals.

To investigate the specific case of *Nature* and *Science*, two comparisons were performed, (i) papers in *Nature* and *Science* that are retrieved from a search on the topic "gene*" with papers published in *Genome Research*, and (ii) papers in *Nature* and *Science* that are retrieved from a search using the topic "material*" with papers published in *Advanced Materials* (the asterisk indicates a truncated word). As with cases already cited in Section 2, the distribution of citations of the *Nature* and *Science* papers published in the selected topics did not follow lognormal distributions—the deviations of the means cited in Section 2.1 were higher than 20%. Therefore, the search was restricted to those papers classified as "articles" by the database. Although the high deviations cited did not occur in either *Genome Research* or *Advanced Materials*, for consistency in all searches, the searches in these journals were also restricted to "articles" only. Table 4 summarizes the characteristics of the citation distributions in the two comparisons, along with the failure probabilities.

Table 4. Failure probability of considering higher merits to papers in *Nature* or *Science* than to papers in specialized journals. Calculations based on lognormal distribution of citations.

| Journal and topic[a] | Number of papers | Mean number of citations | $\alpha$ | $\sigma$ | Failure probability[b] |
|---|---|---|---|---|---|
| *Nature* and *Science* Topic = gene* | 695 | 217.8 | 4.84 | 0.96 | 0.22 |
| *Genome Research* | 238 | 77.0 | 3.75 | 1.05 | |
| *Nature* and *Science* Topic = material* | 111 | 177.9 | 4.65 | 1.01 | 0.24 |
| *Advanced materials* | 805 | 66.3 | 3.67 | 1.04 | |

[a] Citations in 2014-2017 to publications in 2012. Search restricted to "articles"
[b] *Nature* and *Science* versus *Genome Research* and *Nature* and *Science* versus *Advanced Materials*

Although the JIFs are not available in the conditions of the searches, the mean numbers of citations reflect the JIF difference between *Nature* and *Science* and the specialized journals: 217 versus 77, and 178 versus 66 for genes and material topics, respectively. However, despite these great differences, the failure probabilities were high: 0.22 and 0.24, respectively.

## 4. Discussion

*4.1. Citations and "quality"*

The contribution that a paper has had in the progress of its field of research by producing new scientific knowledge is of prime importance in research assessment. This "scientific relevance" or "quality" of a paper is somehow related to its "impact" and citation counts (Leydesdorff et al, 2016). Although the practice of citing (Tahamtan & Bornmann, 2018) and the use of citation counts to measure this or any other effect related to it (e.g., Leydesdorff et al, 2016) still have complexities that are not yet completely clear, citation counts are widely used in scientific evaluations as an indicator of research goodness, either directly or after some kind of normalization as in top percentile indicators (Bormann & Marx, 2013). However, even under this view, it is worth noting that the number of citations correlates but does not measure the scientific relevance of a paper (Section 4.4).

The JIF is a mean of the number of citations of the articles published in a journal, which is calculated for specific publication and citation windows. This indicator of journal's output is in many cases used in substitution of citation counts when they are taken as primary measures of the "scientific relevance". Although this use of JIFs has been widely criticized (Section 1), Waltman & Traag (2017) describe an alternative view that they call scenario 2, in which the scientific merit of a paper could be better described by the JIF than by the number of citations The basis of this scenario is that "journals are fairly homogeneous in terms of the values of the articles they publish" (p. 14) because "the peer review system of a journal will ensure that all or almost all articles in a journal have a value above a certain journal-specific minimum threshold" and "researchers will generally try to publish their work in a journal that is as 'prestigious' as possible" (p. 16). While this alternative view of JIFs could holds for research areas where progress is not rapid, it seems unlikely that it holds in the case of areas of rapid scientific or technological progress (e.g., graphene, Li-batteries, solar cells, cancer, neurodegenerative diseases, etc.). In these latter areas compelling evidence suggests that prestigious journals publish papers of very different scientific or technological relevance, all of which are most certainly above a minimum threshold. However, this or any other discussion about the Waltman & Traag's (2017) scenario 2, which is currently a relevant discussion, needs to be substantiated by surveys among experts.

*4.2. JIFs must not be used as a substitute for citation analyses*

The goodness of any process of decision making has to be judged by its probability of failure; that is, when the process leads to taking a wrong decision. When the JIF is used as a surrogate of a research merit based on citation frequency (e.g., Garfield, 2001; Abramo, D'Angelo, & Di Costa, 2010), the method assigns the same merit to all papers. This use is intuitively misleading because it is impossible that all the papers in a journal have the same number of citations, which is known as "ecological fallacy" (Leydesdorff, Wouters, & Bornmann, 2016). Consequently, many decisions based on the assumption of higher citation merit for papers published in journals with higher JIFs will be wrong. This failure occurs when a paper in a journal with a lower JIF receives more citations than another paper in a journal with a higher JIF and, despite this, it receives a lower evaluation.

It has to be admitted that in specific circumstances, for example, in recently published papers, the use of the JIFs might be reasonable if the risk of making a wrong decision is low. In fact, all processes of decision making carry a risk of failure, including expert reviews in research evaluations. Unfortunately, in many cases, this risk of failure cannot be calculated, and the goodness of a process has to be discussed only by reasoning, which leads to uncertain and always debatable conclusions. This is not the case in our comparison of JIFs with citation counts in evaluations, because in this case the probability of failure can be mathematically calculated, as described in Section 2. Based on this probability we find that the use of JIFs as surrogates of citation frequencies is not suitable.

The results reported in Table 2 show that the failure probability varies depending on the magnitude of the difference between the JIFs that are considered and that the failure probability is high when this difference is low. In many real evaluations, JIFs are used to distinguish between two or more candidates who work in similar research areas and that have similar scientific levels; consequently, the JIFs of the journals where they have published their papers are not very different. Small differences in the JIFs also occur when publications in Q1 journals are evaluated above those published in Q2 journals. In this dichotomous procedure, which is problematic in itself (DeCoster et al, 2009), it can be guessed that two publications, one in a journal in the lower part of the Q1 set of publications and the other in the upper part of the Q2 set will be very similar. Furthermore, in practice, even for publications in journals in the upper part of Q1 versus those in journals in the lower part of Q2, the difference between the JIFs is not high enough to lead to a low probability of failing (Table 3).

All this implies that, in most cases of evaluations, the JIFs are not very different, and failure probabilities are very high (Tables 2 and 3): slightly below 0.5, which is the failure (or success) probability of coin flipping. Tregoning (2018) defends the use of the JIF, claiming that "papers published in journals with higher impact factors tend, on average, to be better and more important than those in journals with lower ones." This is true, and the scientific relevance of the papers published in high-JIF journals is not questioned. The mistake arises when the merits of two papers are judged by the JIF. Aside from doubts about whether scientists can be evaluated by the average of the merits of the papers published by others, the mathematical calculations in our study demonstrate that, in many cases, the risk of failure is unacceptably high. In many cases, it can be as high as in coin flipping, which dismantles all claims based on the tendency of the average.

Another relevant fact that explains the popularity of the use of the JIF for researchers' evaluations is that the number of citations of approximately 70% of the papers published in a journal is below the mean (Rodríguez-Navarro & Brito, 2018b and references therein). This fact implies that for those who believe that citation counts is a measure of merit, the use of the JIF benefits to 70% of the papers. Taking together that (i) for those who distribute money and positions "it is easier to tot up some figures than to think seriously about what a person has achieved" (Lawrence, 2003 p. 259), (ii) the process "is also quick—scanning a list of journals takes very little time—and deeply ingrained" (Tregoning, 2018), and (iii) in comparisons with the number of citations, the JIF over-evaluates 70% of the authors, it can be concluded that the use of the JIF might continue for a long time.

In many countries and institutions, researchers suffer an unbearable pressure to publish in one or two leading journals in their research field, making the journal more important than the scientific message (Lawrence, 2003). This procedure has much in common with the use of JIFs because it assigns the same merit to all publications in the blessed journal. Publications in *Nature* and *Science* occupy the zenith of this policy. Therefore, we calculated the failure probability when papers in *Nature* and *Science* in the topics of genes and materials are compared with papers in two other specific journals that cover the same topics, *Genome Research* and *Advanced Materials* (Section 3.3). Despite the great differences in the mean numbers of citations between papers in *Nature* and *Science* and in the two other journals (≈ 200 versus ≈ 70, respectively; Table 4), the failure probability of assigning more citation merit to *Nature* and *Science* papers is approximately 0.2. This probability is again too high to be acceptable.

If the failure probability in comparative evaluations of two papers by JIFs is high, e.g., 0.5–0.3, it might be considered that when the evaluation is based on several papers of two authors, the risk of making a wrong decision could decrease to a reasonably low level. Although this conclusion could be true, the analysis of this case is complex because it is not clear how to compare two sets of several papers and the result depends on the method of comparison. However, this analysis is of little relevance because the important conclusion from this paper's results—that the JIF should not be used as a surrogate of citation frequency—is methodological and affects to the basis of the method. Furthermore, in real evaluations, this problem is only the tip of an iceberg as shown in the next section.

*4.3. From mathematics to real evaluations*

Our approach calculates failure probabilities by a mathematical technique based on an ideal case, which assumes (i) that the number of citations of a paper reveals its scientific merit and (ii) that papers published in different journals respond exclusively to the merits of the papers without influences from research fields. The first assumption exceeds any consideration about the use of JIFs because it affects to the use of bibliometrics and will be treated in the next section. Regarding the second assumption (ii), it is fulfilled in very few cases. Real evaluations are more complex because the probability of citation of a paper depends not only on its merit but also on the research field (Waltman, 2016). To overcome this problem, evaluations are performed within lists of journals that have been grouped by research areas: for example, the *JCR* categories are used in many evaluations by JIF. The problem is that these and other journal lists correct large differences between the JIFs of the journals that are compared, but the differences that remain are still too large. In other words, papers in different fields within the same area and even in different topics within the same field have different probabilities of reaching a certain citation level (Schubert & Braun, 1986, 1996).

Just to give an example, the *JCR* category of Dentistry, Oral Surgery & Medicine was selected. The first journal in this category is *Periodontology 2000* (2017 JIF = 6.22) and the journal in sixth position is *Journal of Clinical Periodontology* (2017 JIF = 4.05); these two journals cover the dental specialty of periodontics. In contrast, the first journal that specifically covers the specialty of orthodontics and dentofacial orthopedics: *Orthodontics & Craniofacial Research* (2017 JIF = 2.08) is in the 26$^{th}$ position of a total of 91 journals, so it belongs to Q2. Thus, the best publications of top researchers in orthodontics will be in Q2 journals.

Consistent with the difference in the JIFs of the top journals in different specialties, the highest probability that a paper from 2012 received 50 citations up to 2018 (16 October) is 0.058 in the specialty of periodontitis and of 0.008 in the specialty of orthodontics. This large difference between the citation probabilities in these two journals that belong to the same research category demonstrates that the research field penalizes some journals in evaluations by the JIF. The consequence of this issue is that if a faculty or college of dentistry selects its academics by the JIF in the *JCR* category of dentistry or even by citation counting, in a few years it will end up having many academics working in dental infections and no one working in orthodontics.

In summary, probably few evaluations by the JIF are made in research areas that do not include journals with different citation probabilities due to their field scopes. This drawback has to be added to this paper's findings of high failure probabilities that occur in the ideal case, where the probability of citation in all journals depends exclusively on the merit of the papers.

*4.4. Will there be a bibliometric indicator for evaluating researchers?*

In the previous section it is explained that the mathematical calculations of probabilities in this study are based on an ideal case, which assumes that the number of citations of a paper reveals its scientific merit. However, this is not strictly true because scientific relevance and number of citations are only correlated. This notion is behind a large number of publications (e.g., De Bellis, 2009) but it is difficult to study because it requires a numerical scale for the scientific relevance; the best support of this correlation is found in studies that compare citation counts and peer review assessments (Traag & Waltman, 2018; Wilsdon et al, 2015).

A correlation implies that if the regression line is drawn across the data points, most of them will be either above or below the regression line, and many might be far above or below this line. It also implies that if 100 points are chosen at random in the lower, medium, and upper parts of the scatter plot, the average numbers of citations of these sets of data points versus their average scientific relevance will be in the regression line. The conclusion of this simple statistical reflection is that, in research assessment, the use of indicators based on citations has statistical support when the papers of many researchers are aggregated. However, this statistical support does not exist when the approach is applied to an individual paper, an individual researcher, or a low number of researchers. Consequently, a "like-for-like replacement" of the number of citations for the scientific merit assessed by experts cannot be made (Wilsdon et al, 2015, p. ix).

Even assuming ideal cases in which citations measure merit and the lists of journals include a single research field, the use of the JIFs for the evaluation of researchers introduces a bias against novelty because the JIFs are based on a short-term citation window (Wang, Veugelers, & Stephan, 2017) and the citation of these novel papers is delayed. Furthermore, novel papers are not frequently published in journals with the highest JIFs (Wang Veugelers, & Stephan, 2017).

In summary, this mathematical-based study demonstrates that even in ideal lists of journals that do not mix field scopes, the use of JIFs as surrogates of citation frequencies leads to a high proportion of wrong decisions in research evaluations.

Furthermore, if the link between scientific relevance and number of citations is a correlation, the conclusion is that for evaluating individual researchers, no type of citation-based indicators should be used; peer evaluation seems to be the only reasonable alternative.

**Appendix**

*Probability calculations*

Failure probabilities were calculated applying the formula described below. Although citations are integers, the continuous variant of the lognormal distribution was used for the failure probability calculations, which is a reasonable approach (Thelwall, 2016). If, in the case of integers, papers with the same numbers of citations are included as failures in the evaluation by the JIF, the discretized variant of the lognormal distribution will provide a slightly lower probability of failure than the continuous variant. However, this difference is very small; for example, in the two journals *Water Research*, JIF = 4.66, and *Environmental Toxicology and Pharmacology*, JIF = 2.01, the probabilities are 0.18 and 0.17, for the continuous and discretized distributions, respectively.

Let us consider two papers, A and B, published in two different journals called $J_a$ and $J_b$ respectively. We assume that the distributions of citations obey lognormal distributions $p_a(C_a)$ and $p_b(C_b)$, that depend on parameters $\mu_a, \sigma_a$, and $\mu_b, \sigma_b$, respectively. The mathematical expression of the lognormal probability distribution of obtaining $C$ citations is:

$$p(C) = \frac{1}{\sqrt{2\pi}C\sigma} \, exp\left[-\frac{(\ln C - \mu)^2}{2\sigma^2}\right]$$

The goal of this appendix is to calculate the probability that paper B receives more citations than paper A. Start with the probability that A receives $C_a$ citations which is simply given by $p_a(C_a)$. Then the probability that B receives an *equal or greater* number of citations than A is:

$$P_b(C_b > C_a) = p_b(C_a) + p_b(C_a + 1) + p_b(C_a + 2) + \cdots \cong \int_{C_a}^{\infty} dC_b \, p_b(C_b)$$

where it is assumed that a paper B can receive an arbitrarily large number of citations, and, hence, the sum goes to ∞. The joint probability that publication A receives $C_a$ citations and B receives more than $C_a$ citations is the product of both expressions as the events are statistically independent:

$$P(C_a, C_b > C_a) = p_a(C_a) \int_{C_a}^{\infty} dC_b \, p_b(C_b)$$

Finally, to find the failure probability, which is the probability that B receives more citations than A, regardless of the number of citations of A, it is necessary to add all possible values of $C_a$, from zero citations to an arbitrary number:

$$P = \int_0^\infty dC_a\, p_a(C_a) \int_{C_a}^\infty dC_b\, p_b(C_b)$$

The last equation has been used to evaluate the probabilities displayed in Tables 2, 3, and 4.

## Acknowledgements

We would like to thank three anonymous reviewers for their helpful comments, which were used to prepare an improved version of the original manuscript. This work was supported by the Spanish Ministerio de Economía y Competitividad, grant numbers FIS2014-52486-R and FIS2017-83709-R.

## Supplementary material

Additional information may be found in the online version of this article

Table S1. Complete matrix of failure probabilities resulting from pairwise comparisons of 39 journals

Supplementary Material: Failure probability when evaluating by Journal of Impact Factor instead of by the citation counts. See main text for journal details.

| # | 1 | 2 | 3 | 4 | 5 | 6 | 7 | 8 | 9 | 10 | 11 | 12 | 13 | 14 | 15 | 16 | 17 | 18 | 19 | 20 | 21 | 22 | 23 | 24 | 25 | 26 | 27 | 28 | 29 | 30 | 31 | 32 | 33 | 34 | 35 | 36 | 37 | 38 | 39 |
|---|---|---|---|---|---|---|---|---|---|---|---|---|---|---|---|---|---|---|---|---|---|---|---|---|---|---|---|---|---|---|---|---|---|---|---|---|---|---|---|
| 1 | 0.50 | 0.49 | 0.28 | 0.29 | 0.39 | 0.26 | 0.20 | 0.24 | 0.20 | 0.18 | 0.31 | 0.22 | 0.16 | 0.16 | 0.10 | 0.19 | 0.17 | 0.21 | 0.13 | 0.17 | 0.21 | 0.15 | 0.13 | 0.20 | 0.13 | 0.11 | 0.15 | 0.09 | 0.17 | 0.16 | 0.10 | 0.09 | 0.10 | 0.09 | 0.20 | 0.14 | 0.08 | 0.04 | 0.05 |
| 2 | | 0.50 | 0.29 | 0.30 | 0.41 | 0.26 | 0.21 | 0.25 | 0.21 | 0.19 | 0.32 | 0.23 | 0.17 | 0.16 | 0.10 | 0.20 | 0.18 | 0.22 | 0.14 | 0.17 | 0.22 | 0.16 | 0.14 | 0.20 | 0.14 | 0.11 | 0.15 | 0.09 | 0.18 | 0.16 | 0.10 | 0.10 | 0.10 | 0.10 | 0.21 | 0.15 | 0.08 | 0.04 | 0.05 |
| 3 | | | 0.50 | 0.52 | 0.65 | 0.46 | 0.39 | 0.47 | 0.42 | 0.35 | 0.55 | 0.43 | 0.35 | 0.33 | 0.23 | 0.39 | 0.36 | 0.40 | 0.30 | 0.36 | 0.42 | 0.33 | 0.28 | 0.40 | 0.31 | 0.25 | 0.30 | 0.21 | 0.36 | 0.35 | 0.23 | 0.22 | 0.23 | 0.23 | 0.38 | 0.30 | 0.19 | 0.12 | 0.14 |
| 4 | | | | 0.50 | 0.63 | 0.44 | 0.37 | 0.44 | 0.39 | 0.33 | 0.53 | 0.40 | 0.32 | 0.31 | 0.20 | 0.36 | 0.34 | 0.38 | 0.27 | 0.33 | 0.39 | 0.31 | 0.26 | 0.38 | 0.28 | 0.23 | 0.28 | 0.19 | 0.33 | 0.32 | 0.20 | 0.19 | 0.20 | 0.20 | 0.36 | 0.28 | 0.17 | 0.10 | 0.12 |
| 5 | | | | | 0.50 | 0.32 | 0.25 | 0.30 | 0.26 | 0.22 | 0.39 | 0.28 | 0.20 | 0.19 | 0.11 | 0.24 | 0.21 | 0.27 | 0.16 | 0.21 | 0.26 | 0.19 | 0.16 | 0.25 | 0.16 | 0.13 | 0.18 | 0.10 | 0.21 | 0.20 | 0.11 | 0.11 | 0.11 | 0.11 | 0.25 | 0.18 | 0.09 | 0.04 | 0.05 |
| 6 | | | | | | 0.50 | 0.43 | 0.51 | 0.46 | 0.39 | 0.59 | 0.47 | 0.39 | 0.38 | 0.26 | 0.43 | 0.41 | 0.44 | 0.34 | 0.40 | 0.46 | 0.38 | 0.32 | 0.45 | 0.36 | 0.29 | 0.34 | 0.25 | 0.40 | 0.39 | 0.27 | 0.26 | 0.27 | 0.27 | 0.42 | 0.34 | 0.23 | 0.15 | 0.17 |
| 7 | | | | | | | 0.50 | 0.59 | 0.54 | 0.46 | 0.66 | 0.54 | 0.47 | 0.45 | 0.33 | 0.51 | 0.49 | 0.51 | 0.41 | 0.48 | 0.54 | 0.45 | 0.39 | 0.52 | 0.43 | 0.36 | 0.41 | 0.32 | 0.47 | 0.47 | 0.34 | 0.32 | 0.34 | 0.34 | 0.49 | 0.42 | 0.29 | 0.20 | 0.22 |
| 8 | | | | | | | | 0.50 | 0.45 | 0.37 | 0.59 | 0.46 | 0.37 | 0.35 | 0.23 | 0.42 | 0.39 | 0.43 | 0.31 | 0.38 | 0.45 | 0.35 | 0.29 | 0.43 | 0.33 | 0.26 | 0.32 | 0.22 | 0.38 | 0.37 | 0.24 | 0.23 | 0.24 | 0.23 | 0.40 | 0.32 | 0.19 | 0.12 | 0.13 |
| 9 | | | | | | | | | 0.50 | 0.41 | 0.64 | 0.50 | 0.41 | 0.40 | 0.27 | 0.46 | 0.43 | 0.47 | 0.35 | 0.42 | 0.50 | 0.40 | 0.33 | 0.48 | 0.37 | 0.30 | 0.36 | 0.25 | 0.42 | 0.41 | 0.27 | 0.26 | 0.27 | 0.27 | 0.44 | 0.36 | 0.22 | 0.14 | 0.16 |
| 10 | | | | | | | | | | 0.50 | 0.70 | 0.58 | 0.51 | 0.49 | 0.37 | 0.55 | 0.53 | 0.55 | 0.46 | 0.52 | 0.58 | 0.49 | 0.43 | 0.56 | 0.48 | 0.41 | 0.45 | 0.36 | 0.52 | 0.51 | 0.38 | 0.36 | 0.38 | 0.38 | 0.53 | 0.45 | 0.32 | 0.24 | 0.26 |
| 11 | | | | | | | | | | | 0.50 | 0.37 | 0.29 | 0.28 | 0.17 | 0.33 | 0.30 | 0.35 | 0.24 | 0.30 | 0.36 | 0.27 | 0.23 | 0.34 | 0.25 | 0.20 | 0.25 | 0.16 | 0.30 | 0.28 | 0.18 | 0.17 | 0.18 | 0.17 | 0.33 | 0.25 | 0.14 | 0.08 | 0.09 |
| 12 | | | | | | | | | | | | 0.50 | 0.42 | 0.40 | 0.28 | 0.46 | 0.44 | 0.47 | 0.36 | 0.43 | 0.49 | 0.40 | 0.34 | 0.48 | 0.38 | 0.31 | 0.36 | 0.27 | 0.43 | 0.42 | 0.29 | 0.27 | 0.29 | 0.28 | 0.45 | 0.37 | 0.24 | 0.16 | 0.18 |
| 13 | | | | | | | | | | | | | 0.50 | 0.48 | 0.35 | 0.55 | 0.53 | 0.55 | 0.44 | 0.51 | 0.58 | 0.48 | 0.41 | 0.56 | 0.47 | 0.39 | 0.43 | 0.33 | 0.51 | 0.50 | 0.35 | 0.34 | 0.36 | 0.35 | 0.52 | 0.44 | 0.30 | 0.20 | 0.23 |
| 14 | | | | | | | | | | | | | | 0.50 | 0.36 | 0.56 | 0.54 | 0.56 | 0.46 | 0.53 | 0.60 | 0.50 | 0.43 | 0.58 | 0.48 | 0.40 | 0.45 | 0.35 | 0.52 | 0.52 | 0.37 | 0.35 | 0.37 | 0.37 | 0.54 | 0.46 | 0.31 | 0.22 | 0.24 |
| 15 | | | | | | | | | | | | | | | 0.50 | 0.70 | 0.68 | 0.68 | 0.60 | 0.67 | 0.72 | 0.64 | 0.56 | 0.71 | 0.63 | 0.55 | 0.57 | 0.49 | 0.66 | 0.66 | 0.51 | 0.49 | 0.51 | 0.52 | 0.66 | 0.59 | 0.45 | 0.34 | 0.37 |
| 16 | | | | | | | | | | | | | | | | 0.50 | 0.47 | 0.50 | 0.39 | 0.46 | 0.53 | 0.43 | 0.37 | 0.51 | 0.41 | 0.34 | 0.39 | 0.29 | 0.46 | 0.45 | 0.31 | 0.29 | 0.31 | 0.31 | 0.48 | 0.40 | 0.26 | 0.17 | 0.19 |
| 17 | | | | | | | | | | | | | | | | | 0.50 | 0.53 | 0.41 | 0.49 | 0.56 | 0.46 | 0.38 | 0.54 | 0.44 | 0.36 | 0.41 | 0.30 | 0.48 | 0.48 | 0.32 | 0.31 | 0.33 | 0.32 | 0.50 | 0.42 | 0.27 | 0.18 | 0.20 |
| 18 | | | | | | | | | | | | | | | | | | 0.50 | 0.40 | 0.46 | 0.53 | 0.44 | 0.38 | 0.51 | 0.42 | 0.35 | 0.40 | 0.31 | 0.46 | 0.45 | 0.32 | 0.31 | 0.33 | 0.32 | 0.48 | 0.40 | 0.28 | 0.19 | 0.21 |
| 19 | | | | | | | | | | | | | | | | | | | 0.50 | 0.57 | 0.64 | 0.54 | 0.46 | 0.62 | 0.53 | 0.44 | 0.48 | 0.38 | 0.57 | 0.56 | 0.40 | 0.38 | 0.41 | 0.41 | 0.57 | 0.49 | 0.34 | 0.24 | 0.27 |
| 20 | | | | | | | | | | | | | | | | | | | | 0.50 | 0.57 | 0.47 | 0.40 | 0.55 | 0.45 | 0.37 | 0.42 | 0.32 | 0.50 | 0.49 | 0.34 | 0.32 | 0.34 | 0.34 | 0.51 | 0.43 | 0.29 | 0.19 | 0.21 |
| 21 | | | | | | | | | | | | | | | | | | | | | 0.50 | 0.40 | 0.34 | 0.48 | 0.38 | 0.31 | 0.36 | 0.26 | 0.43 | 0.42 | 0.28 | 0.27 | 0.28 | 0.28 | 0.45 | 0.37 | 0.23 | 0.15 | 0.17 |
| 22 | | | | | | | | | | | | | | | | | | | | | | 0.50 | 0.43 | 0.58 | 0.48 | 0.40 | 0.45 | 0.35 | 0.52 | 0.52 | 0.37 | 0.35 | 0.37 | 0.37 | 0.54 | 0.46 | 0.31 | 0.21 | 0.24 |
| 23 | | | | | | | | | | | | | | | | | | | | | | | 0.50 | 0.65 | 0.56 | 0.48 | 0.52 | 0.43 | 0.60 | 0.59 | 0.45 | 0.43 | 0.45 | 0.45 | 0.60 | 0.53 | 0.39 | 0.30 | 0.32 |
| 24 | | | | | | | | | | | | | | | | | | | | | | | | 0.50 | 0.40 | 0.33 | 0.38 | 0.28 | 0.45 | 0.44 | 0.30 | 0.28 | 0.30 | 0.29 | 0.47 | 0.38 | 0.25 | 0.16 | 0.18 |
| 25 | | | | | | | | | | | | | | | | | | | | | | | | | 0.50 | 0.41 | 0.46 | 0.35 | 0.54 | 0.54 | 0.37 | 0.36 | 0.38 | 0.38 | 0.55 | 0.47 | 0.31 | 0.21 | 0.24 |
| 26 | | | | | | | | | | | | | | | | | | | | | | | | | | 0.50 | 0.54 | 0.44 | 0.62 | 0.62 | 0.46 | 0.44 | 0.47 | 0.47 | 0.62 | 0.55 | 0.40 | 0.30 | 0.32 |
| 27 | | | | | | | | | | | | | | | | | | | | | | | | | | | 0.50 | 0.41 | 0.57 | 0.57 | 0.43 | 0.41 | 0.44 | 0.44 | 0.58 | 0.51 | 0.38 | 0.29 | 0.31 |
| 28 | | | | | | | | | | | | | | | | | | | | | | | | | | | | 0.50 | 0.67 | 0.68 | 0.52 | 0.50 | 0.53 | 0.53 | 0.67 | 0.60 | 0.46 | 0.35 | 0.38 |
| 29 | | | | | | | | | | | | | | | | | | | | | | | | | | | | | 0.50 | 0.49 | 0.35 | 0.33 | 0.35 | 0.35 | 0.51 | 0.43 | 0.29 | 0.20 | 0.22 |
| 30 | | | | | | | | | | | | | | | | | | | | | | | | | | | | | | 0.50 | 0.35 | 0.33 | 0.35 | 0.35 | 0.52 | 0.44 | 0.29 | 0.19 | 0.22 |
| 31 | | | | | | | | | | | | | | | | | | | | | | | | | | | | | | | 0.50 | 0.48 | 0.51 | 0.51 | 0.65 | 0.58 | 0.44 | 0.33 | 0.36 |
| 32 | | | | | | | | | | | | | | | | | | | | | | | | | | | | | | | | 0.50 | 0.53 | 0.53 | 0.67 | 0.60 | 0.46 | 0.36 | 0.39 |
| 33 | | | | | | | | | | | | | | | | | | | | | | | | | | | | | | | | | 0.50 | 0.50 | 0.65 | 0.58 | 0.43 | 0.32 | 0.35 |
| 34 | | | | | | | | | | | | | | | | | | | | | | | | | | | | | | | | | | 0.50 | 0.65 | 0.58 | 0.43 | 0.32 | 0.35 |
| 35 | | | | | | | | | | | | | | | | | | | | | | | | | | | | | | | | | | | 0.50 | 0.43 | 0.30 | 0.21 | 0.23 |
| 36 | | | | | | | | | | | | | | | | | | | | | | | | | | | | | | | | | | | | 0.50 | 0.36 | 0.27 | 0.29 |
| 37 | | | | | | | | | | | | | | | | | | | | | | | | | | | | | | | | | | | | | 0.50 | 0.40 | 0.43 |
| 38 | | | | | | | | | | | | | | | | | | | | | | | | | | | | | | | | | | | | | | 0.50 | 0.53 |
| 39 | | | | | | | | | | | | | | | | | | | | | | | | | | | | | | | | | | | | | | | 0.50 |